\documentclass{aa}

\usepackage[varg]{txfonts}
\usepackage{amssymb}
\usepackage{graphicx}

\newcommand\degree{^{\circ}}
\usepackage{graphicx}	% Including figure files
\usepackage{mathtools}	% Advanced maths commands
\usepackage{amssymb}	% Extra maths symbols
\usepackage{multicol}        % Multi-column entries in tables
\usepackage{bm}		% Bold maths symbols, including upright Greek
\usepackage{pdflscape}	% Landscape pages
\usepackage{xcolor}
\usepackage{placeins}
\usepackage{txfonts}

\begin{document} 

   \title{3D MHD simulations of runaway pulsars in core-collapse supernova remnants}

   \author{D. M.-A.~Meyer\inst{1}, 
          D. F. Torres\inst{1,2,3}
          and 
          Z. Meliani\inst{4}
          \\           
          }

    \institute{Institute of Space Sciences (ICE, CSIC), Campus UAB, Carrer de Can Magrans s/n, 
    08193 Barcelona, Spain\\
    \email{dmameyer.astro@gmail.com}
    \and
    Institut d’Estudis Espacials de Catalunya (IEEC), 08034 Barcelona, Spain
    \and
    Institució Catalana de Recerca i Estudis Avançats (ICREA), 08010 Barcelona, Spain
    \and    
    Laboratoire Univers et Théories, Observatoire de Paris, 
    Université PSL, Université de Paris, CNRS, F-92190 Meudon, France
          }
   \date{}

  \abstract
   {  
   Pulsars are one of the possible final stages in the evolution of massive stars. If a supernova explosion is anisotropic, it can give the pulsar a powerful "kick", propelling it to supersonic speeds. The resulting pulsar wind nebula is significantly reshaped by its interaction with the surrounding medium as the pulsar moves through it. First, the pulsar crosses the supernova remnant, followed by the different layers of circumstellar medium formed during different stages of the progenitor star's evolution.
   }
   {     
   We aim to investigate how the evolutionary history of massive stars shapes the bow shock nebulae of runaway "kicked" pulsars, and how these influences in turn affect the dynamics and non-thermal radio emission of the entire pulsar remnant.
   }
   { 
   We perform three-dimensional magnetohydrodynamic simulations using the PLUTO code to model the pulsar wind nebula generated by a runaway pulsar in the supernova remnant of a red supergiant progenitor, and derive its non-thermal radio emission. 
   }
   { 
   The supernova remnant and the pre-supernova circumstellar medium of the progenitor strongly confine and reshape the pulsar wind nebula of the runaway pulsar, bending its two side jets inwards and giving the nebula an arched shape for an observer perpendicular to the jets and the propagation direction, \textcolor{black}{as observed around PSR J1509–5850 and Gemina}. 
   }
   {
   We perform the first classical 3D model of a pulsar moving inward through its 
   \textcolor{black}{supernova ejecta} and circumstellar medium, 
   inducing a bending of its polar jet that turns into characteristic radio synchrotron signature. 
   The circumstellar medium of young runaway pulsars has a significant influence on the morphology and
   emission of pulsar wind nebulae, \textcolor{black}{whose comprehension requires a detailed understanding} 
   of the evolutionary history of the progenitor star. 
   }


   \maketitle

\section{Introduction}
\label{intro}

Pulsars are rapidly rotating, highly magnetized neutron stars. The most plausible scenario for 
their formation involves core-collapse supernova explosions, which occur at the end of a 
massive star’s life.
In particular, \textcolor{black}{stars with an initial mass in the range of 
$8$$--$$25\, \rm M_{\odot}$}, or those 
with larger initial masses and high metallicity, are more likely to be progenitors of neutron 
stars~\citep{woosley_rvmp_74_2002,Heger_2003ApJ...591..288H}.
Massive stars emit powerful winds throughout their lifetimes, which significantly influence 
their environment by shaping the surroundings and depositing large amounts of stellar 
material and energy into the interstellar medium (ISM) \citep{weaver_apj_218_1977}.
When a massive star explodes as a supernova, the young supernova ejecta propagates through 
the circumstellar medium (CSM) created by the progenitor star~\citep{meyer_mnras_450_2015}.
This interaction decelerates the supernova ejecta, extracting both momentum and energy 
\citep{Dessart_2024arXiv240504259D}, and in some cases, can confine or reshape it 
\citep{meyer_mnras_515_2022}.

If a pulsar forms after a supernova explosion, it generates a pulsar wind nebula (PWN) 
or plerion.
The pulsar wind is extremely powerful and relativistic, reaching up to $10^{38}, \rm erg, \rm s^{-1}$ 
with speeds close to that of light.
This wind first interacts with the core-collapse supernova ejecta, and subsequently with the 
circumstellar medium of the massive progenitor star~\citep{behler_RPPh_2014}.
The interaction between the pulsar wind nebula and the reverse shock of the supernova remnant 
(SNR) initiates the phase known as reverberation (see \cite{Bandiera_mnras_165_2023,2023_mnras_2839_2023}). 
Simulations have shown that the encounter between an anisotropic pulsar wind nebula and the reverse 
shock of the SNR can also induce instabilities and turbulence within the SNR, resulting in 
significant mixing of materials~\citep{orlando_aa_645_2021, orlando_aa_666_2022, meyer_mnras_521_2023}.

\begin{figure*}
        \centering
        \includegraphics[width=0.30\textwidth]{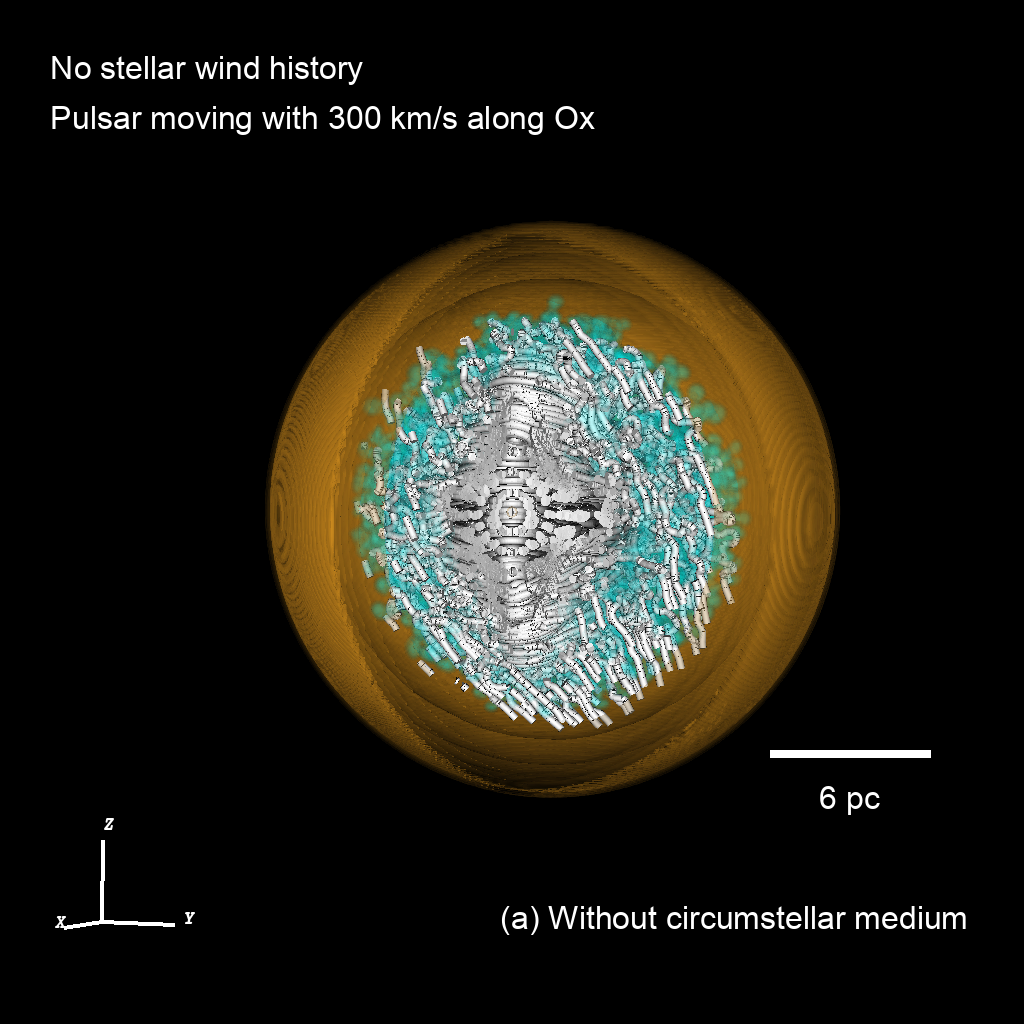}  
        \includegraphics[width=0.30\textwidth]{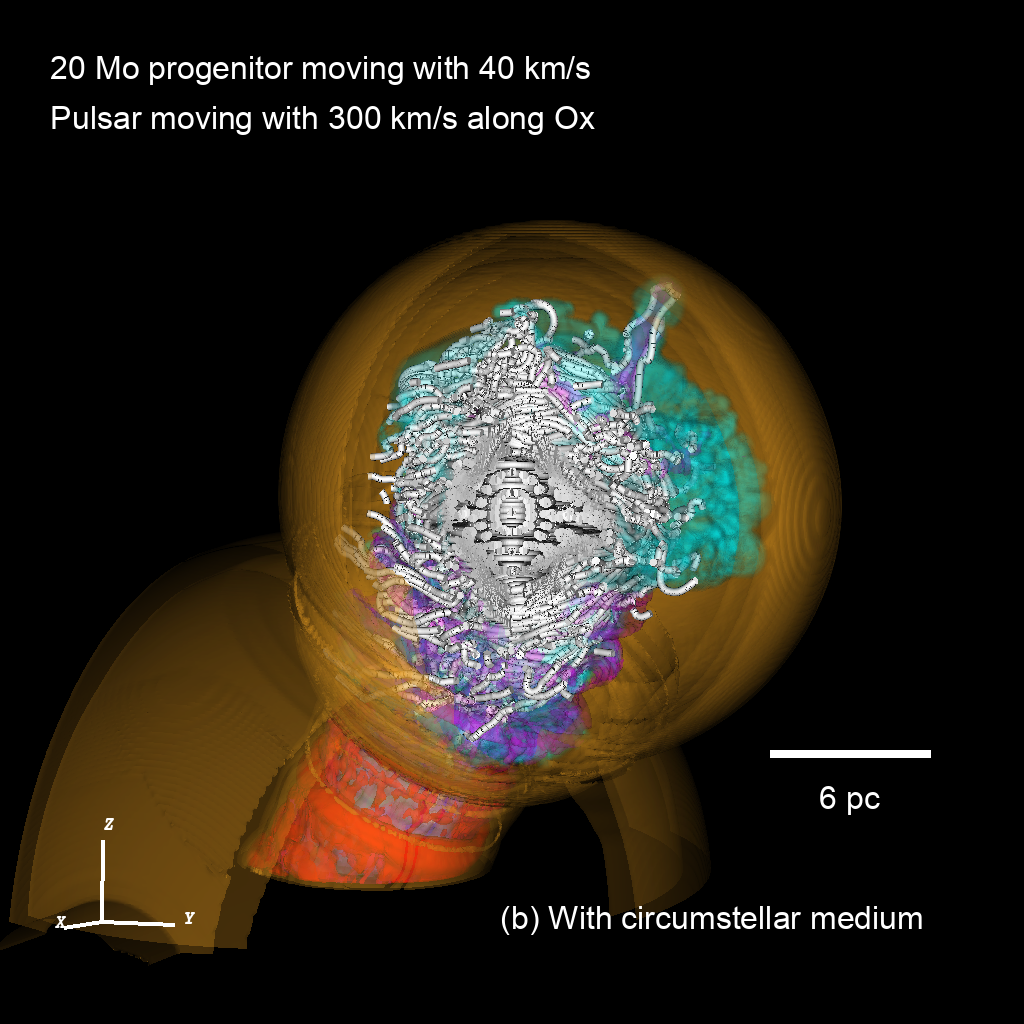}  \\
        \includegraphics[width=0.30\textwidth]{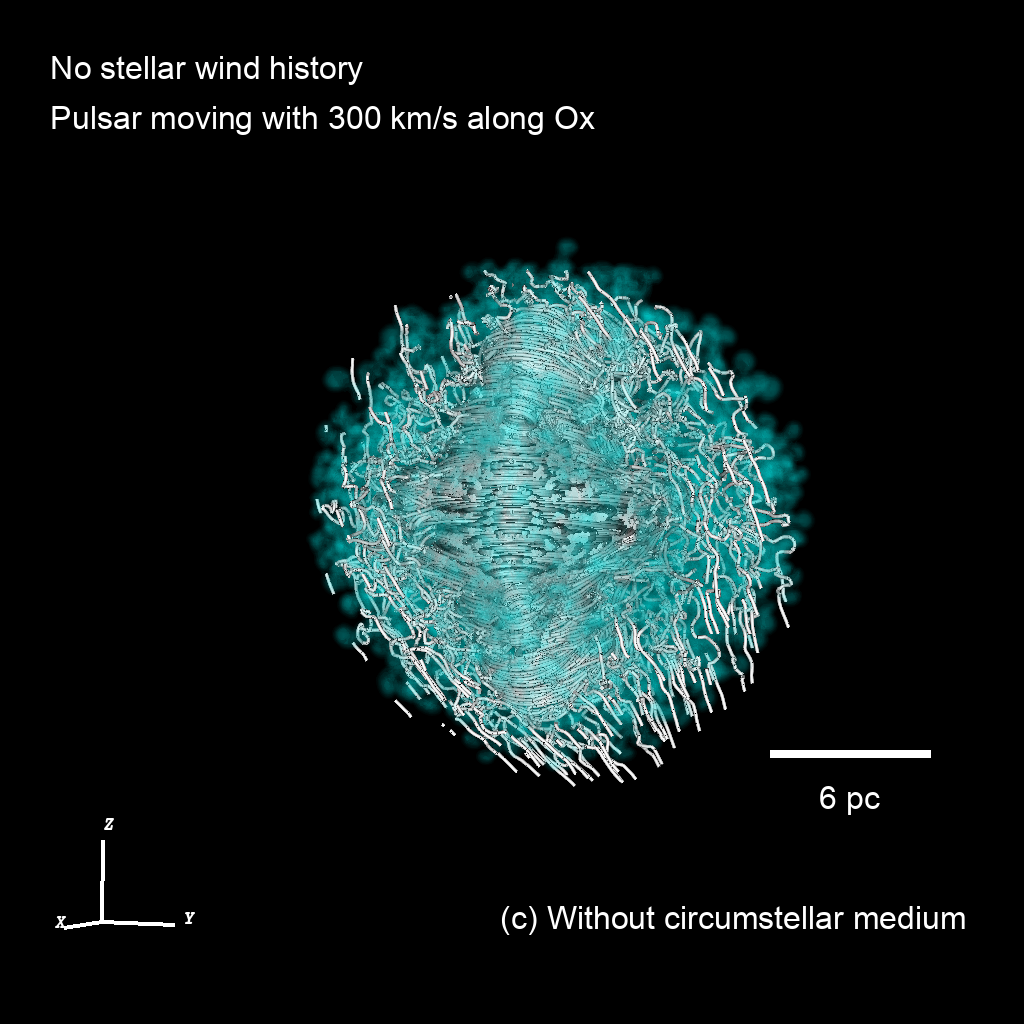}  
        \includegraphics[width=0.30\textwidth]{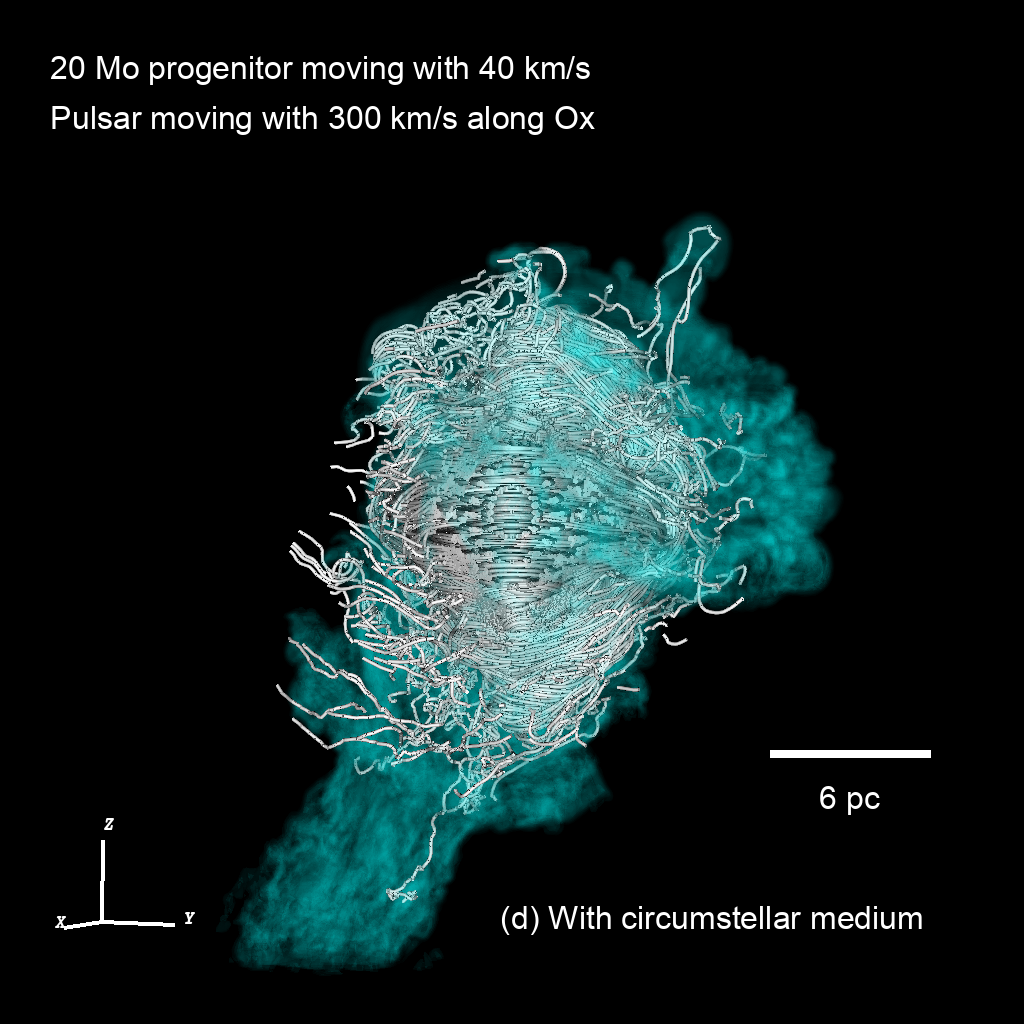}  \\        
        \caption{
        Rendering of the 3D MHD pulsar wind nebula models, considered without (a, c) and 
        with (b, d) and their complex environments. 
        The orange surfaces \textcolor{black}{trace constant density regions of the stellar wind 
        bubble. The remaining} color coding is as follows:
        red traces the supergiant wind,
        blue traces the Wolf-Rayet wind,
        cyan traces the supernova ejecta, and
        white tubes trace the 
        magnetic field lines in the pulsar wind. 
        The 3D figures are cut with two planes 
        permitting a visualization of the internal structure of the plerionic supernova remnants. 
        }
        \label{fig:3D_PWN_rendering}  
\end{figure*}

These regions of multiple shocks and discontinuities, become important sites for cosmic 
ray production through particle acceleration, see e.g, \citep{das_aa_689_2024}, as well as non-thermal 
emission, see e.g., \cite{orlando_aa_622_2019}. Studying plerionic supernova remnants in detail provides 
insights into the history of massive stars and helps constrain stellar evolution models, along 
with the physics of the ISM \citep{langer_araa_50_2012}. 
The geometry of pulsar wind nebulae, \textcolor{black}{at least for young pulsars, is intrinsically 
linked} to the 
morphology of the medium in which the progenitor star evolved and eventually 
died~\citep{blondin_apj_563_2001, bucciantini_an_335_2014, kargaltsev_ssrv_191_2015}. 
This connection becomes particularly significant when the progenitor star was a runaway star 
moving supersonically through the ISM prior to its explosion~\citep{hoogerwerf_apj_544_2000}, 
transforming its pre-supernova CSM into a stellar wind bow shock~\citep{wilkin_459_apj_1996}.
Additionally, the pulsar can receive a birth kick during the explosion, imparting a proper motion 
with velocities of several hundred kilometers per second~\citep{igoshev_mnras_494_2020}. The pulsar 
then moves first through the supernova ejecta, then the CSM, and eventually into the ISM.
As a result, the shape and appearance of PWNe become a complex problem governed by multiple 
physical processes and influenced by a wide range of parameters. 
Understanding this geometry requires considering not only the  motion of the pulsar and 
environment but also the dynamic evolution of the surrounding medium over time.
Recent observations from the {\it James Webb Space Telescope (JWST)} in the infrared offer a 
new opportunity to investigate young supernova remnants  and the environments in which 
young kicked pulsars evolve~\citep{Milisavljevic_2024ApJ...965L..27M}.

In last decades, the community has made extensive efforts to investigate PWNe numerically. 
However, the high complexity of this problem often necessitates numerous simplifying assumptions 
to facilitate analysis. Various models, motivated by the study of the Crab Nebula, have been performed 
in 1D, 2D, and 3D and focusing on the expansion of pulsar wind nebulae into uniform media, \textcolor{black}{  see e.g.,
~\citep{komissarov_mnras_349_2004, Del_Zanna_etal_2006A&A...453..621D, 
2017ASSL..446..215D,
porth_mnras_431_2013,
Olmi_etal_2014MNRAS.438.1518O,Porth_etal_2014MNRAS.438..278P}.}
The consideration of the natal kick received by the pulsar has also generated a body of literature 
involving 2D and 3D simulations that encompass the entire structure where the pulsar wind interacts 
with the ISM, see e.g.,
\textcolor{black}{
~\citep{Bucciantini_aa_375_2001, bucciantini_mnras_478_2018,2019MNRAS.484.5755O}.}
Non-magnetized 3D models have been developed to investigate interactions between the winds of moving 
pulsars and expanding supernova ejecta \citep{temim_apj_808_2015, kolb_apj_844_2017, temim_apj_851_2017,
temim_apj_932_2022}. Additionally, 2D magnetized models for static pulsars that include the progenitor's 
past history have been presented by \citet{meyer_mnras_515_2022}.

In this work, we extend these numerical efforts by presenting a 3D magnetized model for a 
runaway pulsar, which accounts for both the supernova ejecta and the complex CSM of its progenitor.

\begin{figure*}
        \centering
        \includegraphics[width=0.75\textwidth]{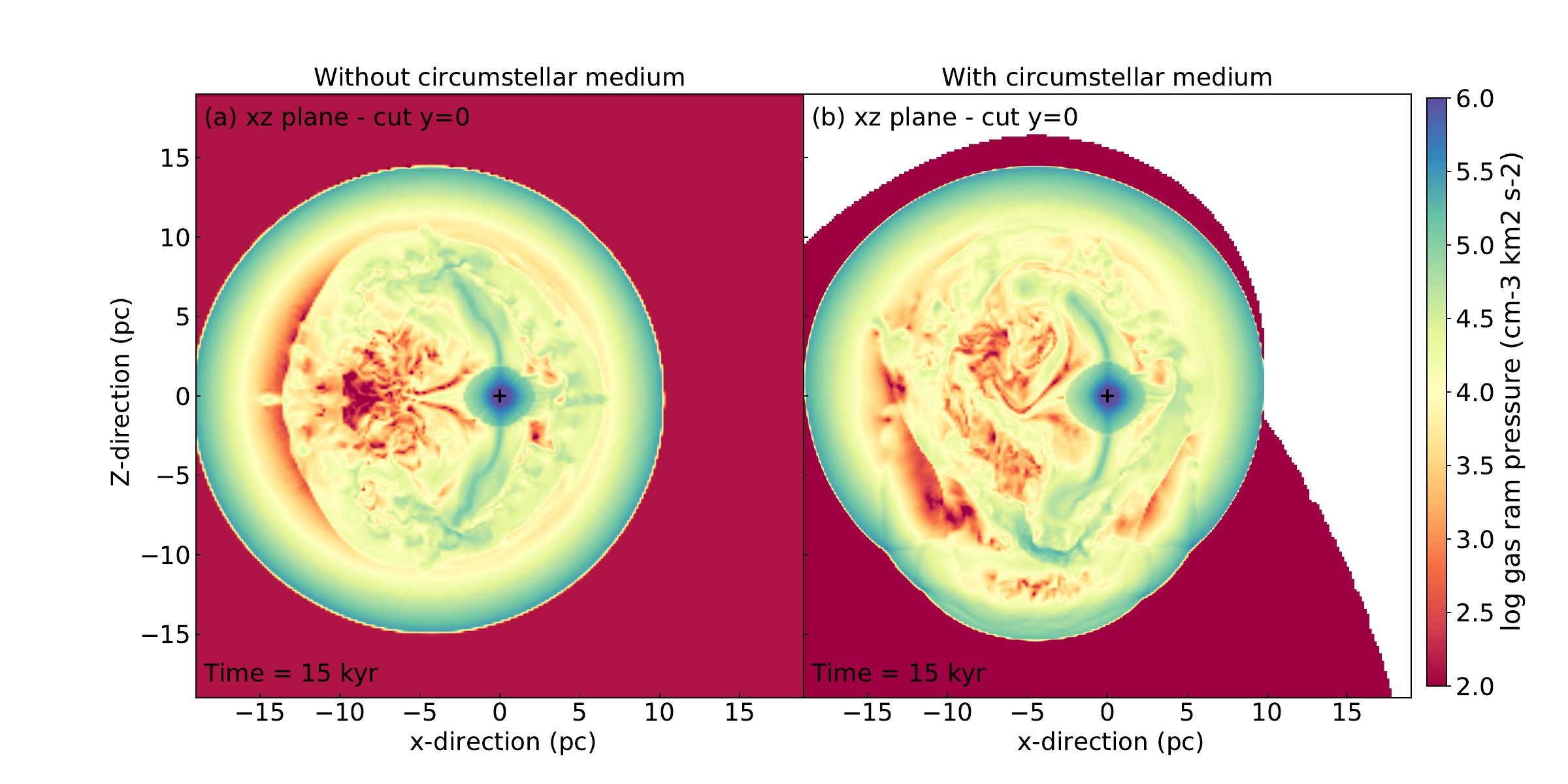}  
        \caption{
        Ram pressure in the $y=0$ plane of the models, without (a) and with (b) circumstellar medium. 
        }
        \label{fig:pressure}  
\end{figure*}

\section{Numerical setup}
\label{method}

A global \textcolor{black}{classical magneto-hydrodynamics} simulation of the CSM 
around a runaway massive 
star in the warm ISM of the Milky Way's galactic plane is performed, following its evolution 
up to the plerionic SNR phase using the {\sc PLUTO} code 
\citep{mignone_apj_170_2007,migmone_apjs_198_2012}. 
\textcolor{black}{
The classical approximation for pulsar wind follows, e.g. 
\citet{blondin_apj_563_2001,temim_apj_851_2017}. 
}

We first investigate a model of a runaway star with an initial mass of $20\, \rm M_{\odot}$, 
\textcolor{black}{which is entering its final evolutionary phase as a red supergiant}. 
The star moves with velocity $v_{\star} = 40\, \rm km\, \rm s^{-1}$ through the warm phase of 
the galactic plane in the Milky Way \citep{blau1993ASPC...35..207B}, where the temperature is 
$8000\, \rm K$ and the number density is $0.79\, \rm cm^{-3}$ \citep{wolfire_apj_587_2003,meyer_2014bb}. 
The magnetic field in the ISM, with a strength of $7\, \mu \rm G$, is assumed to be aligned with the 
star's direction of motion \citep{meyer_mnras_464_2017}. For the thermal processes in the wind nebula, 
we apply optically-thin cooling and heating appropriate for fully ionized gas \citep{wiersma_mnras_393_2009}, assuming solar metallicity \citep{asplund_araa_47_2009}, as described in \citet{meyer_2014bb}.
Stellar surface parameters, including wind velocity, effective temperature, and mass-loss rate as 
a function of time, are derived from the non-rotating zero-age main-sequence model of 
$20\, \rm M_{\odot}$, interpolated from the {\sc Geneva} evolutionary track library  
\citep{ekstroem_aa_537_2012}. Terminal wind velocities are derived using the method 
described in \citet{eldridge_mnras_367_2006}.
The simulation for the CSM surrounding the massive star is conducted in a 
2.5D cylindrical coordinate system (2D plus an additional toroidal component for the 
vectors), utilizing a uniform grid that spans $[0,150] \times [-50,50]\, \rm pc$  
consisting of $7500 \times 5000$ grid zones.

The supernova explosion phase is initially simulated in 1D spherical symmetry, since 
during this initial phase, the SNR propagates through an isotropic medium. Specifically, 
the supernova blast wave is set within the red supergiant stellar wind in 1D. 
For this purpose, we employ a radial grid spanning $[0, 2.0]\, \mathrm{pc}$, 
mapped with $50000$ uniform grid zones. 
We adopt the core-collapse explosion model from \citet{whalen_apj_682_2008} and \citet{truelove_apjs_120_1999}, using an ejecta mass of $M_{\rm ej} = 6.96\, \mathrm{M_{\odot}}$ 
and a canonical explosion energy of $E_{\rm ej} = 1\, \mathrm{foe}$. 
The 1D simulation runs until the supernova blast wave reaches a distance of 
$1.2\, \mathrm{pc}$ from the progenitor star. 
The results from the 1D simulation are then mapped onto a 3D Cartesian coordinate system 
spanning $[-20, 20]^{3}\, \mathrm{pc}$, consisting of $1024^{3}$ grid zones, which is 
pre-filled with the 2.5D CSM assuming rotational symmetry and introducing a $30$ degrees 
shift between the cylindrical axis of symmetry and the Cartesian $Oz$ axis, avoiding  grid effects. 
\textcolor{black}{
We ensure that no information is lost in the wind-ISM interaction when modeling the 
pulsar wind nebula by using a CSM distribution with a higher spatial resolution than 
that of the SNR. The latter is chosen to properly resolve the termination shock of 
the pulsar's bow shock, preventing boundary effects that could introduce numerical 
artifacts, such as those seen in Fig. 2 of \citet{temim_apj_851_2017}. 
}

About $20$ years after the supernova explosion, the pulsar wind is set at the surface of a 
sphere with a radius of $\sim 0.4 \,\rm pc$ (corresponding to $20\, \rm cells$ in the simulation box). 
We follow the prescriptions given by \citet{komissarov_mnras_349_2004} for the pulsar wind. 
Moreover, the pulsar wind is set with a mechanical power luminosity of 
$\dot{E}_{\rm o} = 10^{38}\, \rm erg\, \rm s^{-1}$, a velocity up to $1\%$ of the speed of 
light, and a magnetisation of $\sigma = 10^{-3}$ \citep{2017hsn_book_2159S}. It should be 
noted that in this paper, we use a non-relativistic pulsar wind.
The initial pulsar spin is set to $P_{\rm o} = 0.3\, \rm s$, with a time variation of 
$\dot{P}_{\rm o} = 10^{-17}\, \rm s\, \rm s^{-1}$ and a braking index of $n = 3$ 
\citep{pacini_1973,2017hsn_book_2159S}, following the model used 
in \citet{meyer_mnras_515_2022,meyer_527_mnras_2024}. 
We consider the pulsar’s velocity as $v_{\rm psr} = 300\, \rm km\, \rm s^{-1}$ along the $Ox$ axis,
perpendicularly to the initial direction of motion of the runaway progenitor star, 
also introducing small arbitrary angles ($<$ few degrees) between it and the 
Cartesian characteristic directions, to get rid of grid effects. The simulation grid is set in 
the reference frame of the pulsar, and the ISM is given a velocity of $-v_{\rm psr}$ 
\citep{verbunt_aa_608_2017,igoshev_mnras_494_2020}. 
The simulation parameters are chosen to match those of the most common galactic SNRs 
associated with runaway massive progenitors (see discussion in \citet{meyer_aa_687_2024}). 
Further exploration of this problem is highly desirable, e.g. in terms of 
variations in the initial progenitor star mass, velocity, and supernova energy.

Radiative transfer calculations are performed using the \textsc{radmc-3d} code \citep{dullemond_2012} 
on the results of the MHD simulations. The synthetic radio waveband observables resulting from 
synchrotron emission are generated \citep{meyer_aa_687_2024}. Two simulations are conducted: 
one without and one with the CSM of the core-collapse progenitor prior to the supernova.

\begin{figure*}
        \centering
        \includegraphics[width=0.60\textwidth]{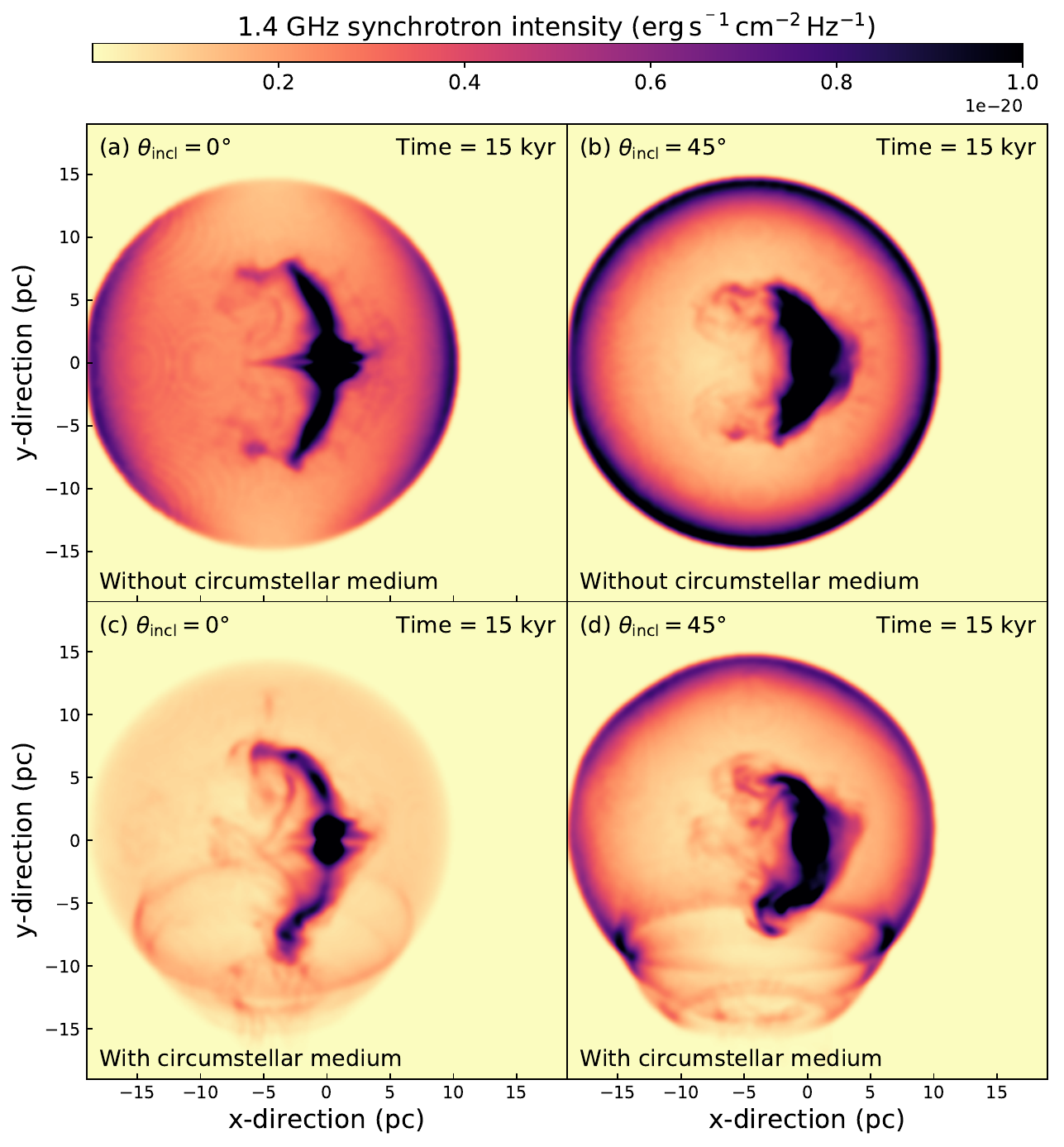}  
        \caption{
        Non-thermal $1.4\, \rm GHz$ radio synchrotron emission maps 
        (in $\rm erg\, \rm s^{-1}\, \rm cm^{-2}\, \rm sr^{-1}\, \rm Hz^{-1}$) 
        for the pulsar wind nebulae and their host supernova remnant, 
        without (a,b) and with (c,d) circumstellar medium. 
        The maps are displayed with a viewing angle of $\theta_{\rm incl}=0\degree$ (a,c) 
        and $\theta_{\rm incl}=45\degree$ (b,d). The dark color marks the regions of 
        strong emission intensity, while the pale color marks the regions of fainter 
        radio emission intensity, respectively.
        }
        \label{fig:3D_PWN_maps}  
\end{figure*}

\section{Results}
\label{results}

\subsection{Without pre-supernova circumstellar medium}
\label{no_csm}

To constrain the influence of the progenitor star’s CSM on the PWN, we first model 
that of a kicked-off pulsar within an isotropic SNR surrounded by a static ISM 
(Fig. \ref{fig:3D_PWN_rendering}a). 
In that figure, the orange surfaces are isodensity contours \textcolor{black}{highlighting 
the forward shock of the blast wave propagating} through the uniform ISM. 
Approximately 15 kyr after the supernova explosion, the pulsar moves through the low-density 
bubble created by the expanding SNR before it begins interacting with the reverse shock and, 
later, with the contact discontinuity of the SNR. 
Due to the anisotropy of the PWN, these interactions with the SNR's reverse shock trigger 
the development of Richtmyer-Meshkov instabilities at the contact discontinuity, as 
described by \citet{kane_apj_511_1999}. 
These instabilities are visible in Figure \ref{fig:3D_PWN_rendering}a, where the SNR tracer 
is shown in cyan.
The magnetic field, illustrated in Figure \ref{fig:3D_PWN_rendering}a with white tubes, 
becomes turbulent in the unstable regions of the SNR. As the 
PWN propagates through the SNR, the magnetic field of the supernova remnant is compressed 
and connects to the structured magnetic field of the PWN. 
The PWN also bends laterally as the pulsar moves within the SNR. 
This bending effect can be observed by an observer situated perpendicular to both the 
jet’s propagation direction and its axis. Indeed, the jet is inclined backward by an 
angle of approximately \textcolor{black}{35 degrees}.

\subsection{With pre-supernova circumstellar medium}
\label{with_csm}

In this simulation, we first allow the CSM of the runaway $20\, \rm M_{\odot}$ 
massive star to evolve until it explodes as a supernova, approximately $15\, \rm kyr$ after the 
explosion, as shown in Figs. \ref{fig:3D_PWN_rendering}b,d,c. 
In these figures, the red surfaces trace the red supergiant wind, while the blue surfaces 
represent the Wolf-Rayet wind. 
The supernova ejecta expands within the CSM and is deformed by it, resulting in the SNR 
adopting a Cygnus-loop-like morphology~\citep{meyer_aa_687_2024}. 
The ejecta forms a strongly elongated region, with an extended cavity to the south and a 
northern outflow that sweeps up the CSM and ISM. 
This behavior aligns with the mechanism described by \citet{2023MNRAS.521.5354M}, 
where post-main-sequence material is channeled into the stellar wind cavity (represented by 
the red and blue contours in Figs. \ref{fig:3D_PWN_rendering}b), which in turn causes 
deviations from sphericity in the expanding supernova ejecta (cyan surface).
The material from the stellar wind becomes confined between the supernova remnant and the PWN, 
which expands and moves laterally with the pulsar. 
The higher density and temperature of the channeled \textcolor{black}{CSM} and 
supernova material, compared 
to the previous simulation, increase the thermal pressure, leading to stronger confinement of 
the PWN's jet. 
This results in more bent and collimated jet than in the first case.

\section{Discussion and conclusion}
\label{discussion}

Using classical magneto-hydrodynamics, 
we simulate the propagation of a kicked-off young pulsar through the CSM of its progenitor 
star for the first time in 3D. 
By incorporating the detailed history of the progenitor star, magnetic field effects, and higher 
spatial resolution, our model provides a more accurate profile of the SNR hydrodynamic simulations, 
such as those by \citet{temim_apj_851_2017} in their modeling of the SNR MSH 15-56.
The anisotropic PWN interacts with the SNR, enhancing the development of instabilities. 
Simultaneously, the SNR and backflow of older CSM, converging toward the 
center of the CSM, further confine the PWN. As the pulsar moves outward, this material bends 
the jet within the nebula at an angle of approximately \textcolor{black}{40$-$50 degrees}. Furthermore, because the 
CSM of the moving progenitor star has a dense bow shock at the front and a lower-density tail, 
the jet propagating toward the tail experiences less bending. As this side of the jet reaches 
the CSM’s tail, it begins to propagate more linearly along the tail’s axis. However, as the 
pulsar continues to move, this side of the jet also gradually bends.

To better quantify the effect of the CSM on the morphology of the PWN, we plot the ram pressure 
$n v^2$ — where $n$ is the gas number density and $v$ the velocity field in the SNR — in the 
frame of reference of the moving pulsar (see Fig. \ref{fig:pressure}). This figure shows 
the ram pressure distribution in the $y=0$ plane, including the pulsar’s motion along the 
$Ox$ axis, with the pulsar located at the origin. The structure of the stellar wind bow shock 
is evident in this figure, along with the high ram pressure regions corresponding to the 
free-streaming pulsar wind and the bent vertical jets. The post-shock region at the forward 
shock of the supernova blastwave interacting with the unperturbed ISM also exhibits significant 
ram pressure, particularly in the southern region of the SNR, where localized areas apply 
pressure on the pulsar's polar jets. These regions result from the reflection of the supernova 
blastwave's termination shock against the CSM, altering the morphology of the PWN as seen in 
Fig. \ref{fig:3D_PWN_rendering}a-d.

Emission maps are calculated using the emission coefficients of \citet{meyer_aa_687_2024}. 
It is assumed that the radiative emission is synchrotron radiation emitted by a population of 
relativistic electrons, with an energy distribution following a power law proportional to 
$\gamma^{-p}$, where $\gamma$ is the Lorentz factor and $p = 2.2$. In our approach, we assume 
that $1\%$ of the gas number density consists of electrons and that $1\%$ of the thermal 
energy of the gas is transferred to the electron population. Our calculations are limited 
to radio frequencies, specifically $\nu = 1.4\, \rm GHz$. Furthermore, for the sake of 
simplicity, we have chosen to neglect the effects of self-absorption and a simplified 
electron energy distribution in the integration of the radiative transfer equation 
along the line of sight. 
These assumptions may influence maps interpretation, particularly 
in dense CSM or high optical depths, as those factors might lead to overestimating 
emission at frequencies much lower than $1.4\, \rm GHz$, as it alters spectral shapes in 
compact PWN \citep{2009ApJ...703.2051G}. 
Additionally, adopting a simplified power-law electron distribution may overlook 
spectral variations observed in real systems \citep{2017hsn_book_2159S}.

The resulting synthetic images were viewed from different angles (with $\theta_{\rm obs} = 0^\circ$ 
and $\theta_{\rm obs} = 45^\circ$ from the $Oy$ axis, which is perpendicular to both the progenitor 
star's motion and the kicked pulsar's). The movement of the CSM is illustrated in Fig. \ref{fig:3D_PWN_maps}a,b for the model without CSM and in Fig. \ref{fig:3D_PWN_maps}c,d for the model with CSM. In both 
models, the emission is dominated by the supernova blast wave and the PWN. It exhibits 
an arc-like morphology with filamentary structures trailing behind the moving pulsar.
The apparent morphology of the PWN is dependent on the observation angle. At an observation 
angle of $0^\circ$, the jet within the PWN appears more elongated and thinner than it does 
at an observation angle of $45^\circ$, where the jet appears broader.
Our study proposes a novel scenario to produce bending of runaway pulsar's tails 
as a direct results of the dense CSM material encountered by the polar material in the 
pulsar wind, naturally explaining north-south asymmetries, such as observed around young 
pulsars still moving in their host remnant, like PSR J1509–5850  \citep{bordas_aa_644L_2020} 
and Gemina  \citep{posselt_apj_835_2017}.

The controlling effect of the runaway pulsar bow shock on the radio appearance of the SNR — governed 
by the pulsar properties (relativistic wind velocity and wind power luminosity) and by its the 
local conditions — provides insights into the age of the SNR, its direction of motion, and the past evolution 
of its progenitor. 
This effect also influences the distribution of the various chemical species involved in the problem, 
namely the ejecta and the different stellar winds blown during the red supergiant and Wolf-Rayet 
evolutionary phases, respectively. 
Our setup can be used in the future to perform models tailored to specific objects.

\begin{acknowledgements} 

The author acknowledges RES resources provided by 
BSC in MareNostrum to \textcolor{black}{AECT-2025-1-0004}. 
The authors also acknowledge computing time 
on the high-performance computer "Lise" at the NHR Center NHR@ZIB. This 
center is jointly supported by the Federal Ministry of Education and 
Research and the state governments participating in the NHR 
(www.nhr-verein.de/unsere-partner). 
This work has been supported by the grant PID2021-124581OB-I00 funded by 
MCIU/AEI/10.13039/501100011033 and 2021SGR00426 of the Generalitat de Catalunya. 
This work was also supported by the Spanish program Unidad de Excelencia Mar\'ia 
de Maeztu CEX2020-001058-M with funding from European Union 
NextGeneration EU (PRTR-C17.I1). 

\end{acknowledgements}

\bibliographystyle{aa} 
\bibliography{grid}

\end{document}